\documentclass[twocolumn,showpacs,preprintnumbers,floats,aps,pra,10pt]{revtex4-1}
\usepackage{graphicx}
\usepackage{dcolumn}
\usepackage{amsmath}
\usepackage{amsfonts}
\usepackage{amssymb}
\usepackage{color,amsxtra}
\usepackage[colorlinks=true,
            linkcolor=blue,
          urlcolor=gray,
            citecolor=red]{hyperref}
\providecommand{\U}[1]{\protect\rule{.1in}{.1in}}

\def\be{\begin{equation}}
\def\ee{\end{equation}}
\def\bea{\begin{eqnarray}}
\def\eea{\end{eqnarray}}

\def\nn{\nonumber \\}

\begin{document}
\title{Analysing dissipative effects in the $\Lambda$CDM model}

\author{Norman Cruz}
\altaffiliation{norman.cruz@usach.cl}
\affiliation{Departamento de F\'isica, Universidad de Santiago de Chile, \\
Avenida Ecuador 3493, Santiago, Chile.}

\author{Esteban Gonz\'alez}
\altaffiliation{esteban.gonzalezb@usach.cl}
\affiliation{Departamento de F\'isica, Universidad de Santiago de Chile, \\
Avenida Ecuador 3493, Santiago, Chile.}

\author{Samuel Lepe}
\altaffiliation{samuel.lepe@pucv.cl}
\affiliation{Instituto de F\'isica, Facultad de Ciencias, \\
Pontificia Universidad Cat\'olica de Valpara\'iso, \\
Avenida Brasil 4950, Valpara\'iso, Chile.}

\author{Diego S\'aez-Chill\'on G\'omez}
\altaffiliation{diego.saez@ehu.eus}
\affiliation{Department of Theoretical Physics, University of the Basque Country UPV/EHU, \\ P.O. Box 644, 48080 Bilbao, Spain}

\date{\today}

\begin{abstract}
In the present paper, the effects of viscous dark matter are analysed within the $\Lambda$CDM model. Here we consider bulk viscosity through the Israel-Stewart theory approach, leading to an effective pressure term in the continuity equation that accomplishes for the dissipative effects of the dark matter fluid. Then, the corresponding equation for viscosity is solved and a general equation for the Hubble parameter is obtained with the presence of a cosmological constant. The existence of de Sitter solutions is discussed, where a wider range of solutions is found in comparison to the $\Lambda$CDM model. Also the conditions for the near thermodynamical equilibrium of the fluid is analysed. Finally, a qualitative analysis provides some constraints on the model by using Supernovae Ia data which reveals the possible importance of causal thermodynamics in cosmology.\end{abstract}
 \maketitle

\section{Introduction}

Late-time acceleration has become one of the main challenges in theoretical physics nowadays as reflects an anormal behaviour of the universe expansion within the Big Bang model, an acceleration that started recently in cosmic time terms. In order to achieve such behaviour, an effective fluid with negative pressure is required, known as dark energy, which in general violates at least one of the energy conditions, and whose equation of state (EoS) must be around -1, according to most of the analysis. In addition, dark energy is the majority component in the universe, representing about three quarters of the total composition of the universe. Despite the available cosmological data coming from Supernovae Ia (SNe Ia)~\cite{acceleration,Suzuki:2011hu}, cosmic microwave background radiation (CMB) \cite{Aghanim:2018eyx}, baryon acoustic oscillations (BAO) \cite{Eisen05} and the recent Hubble data \cite{Hdata} seem to be well described by the so-called $\Lambda$-Cold Dark Matter ($\Lambda$CDM) model, plenty of efforts have been spent on the analysis of different candidates to accomplish such phenomena. From scalar fields, modified gravities to the vacuum energy density with aggressive fine-tunings, many different models have been proposed that are capable of reproducing the late-time acceleration (for a review on dark energy candidates, see Ref.~\cite{Frieman:2008sn,Nojiri:2010wj}). While some dark energy models assume the existence of new fields \cite{Frieman:2008sn}, some others are shown as the incompleteness of our theories (modified gravities) \cite{Nojiri:2010wj}. Nevertheless, nowadays the main issue lies on the difficulties to distinguish among the many existing models that can explain the acceleration of the universe expansion, leading to a problem of degeneracy that is far from being solved. In general, most of these models are constructed in such a way that mimic an effective cosmological constant nowadays, or in other words behaves approximately as  the $\Lambda$CDM model, describing the expansion somewhat close to a de Sitter solution, i.e. an exponential expansion. Hence, one of the main branches being studied today is focused on searching new effects and predictions, specially at the perturbative level from each model, which may provide a way to break down the degeneracy among the models. \\

On the other hand, when studying cosmology, one usually assumes a perfect fluid as an approximated well description of the universe content. However, perfect fluids are descriptions in thermodynamical equilibrium that do not increase the entropy, whose dynamics are reversible. Only when analysing perturbations, non-adiabatic processes are considered. In this sense, dark matter is assumed to behave as a perfect fluid with no pressure, i.e. by a non-relativistic EoS. Nevertheless, in causal thermodynamics, additional ingredientes should be assumed in order to provide a more realistic description, as viscosity \cite{Maartens:1996vi,Israel1979}. Hence, regardless of the nature of dark matter, one may try to infer properties beyond the perfect fluid picture for dark matter by studying the cosmological evolution \cite{Velten:2011bg,Blas:2015tla,Li:2009mf}. In addition, some analysis focus on the study of dissipative effects as an alternative to dark energy, i.e. on the possibility of producing late-time acceleration based on viscous dark matter \cite{Velten:2011bg,Li:2009mf,Padmanabhan:1987dg}, which indirectly means the unification of dark matter and dark energy under a particular form of the EoS \cite{Bruni:2012sn}. Some other previous works focus on the existence of viscous dark energy \cite{Capozziello:2005pa,Nojiri:2004pf,Deffayet:2010qz,Cataldo:2005qh}. There are two types of viscosity that usually arise in hydrodynamics, the shear and bulk viscosity. As the universe seems to be highly isotropic at large scales, the shear viscosity is assumed to be null at least at late-times, despite it may turn out important in some scenarios \cite{Floerchinger:2014jsa}. In addition, bulk viscosity may lead to an effective equation of state that changes along the cosmological history \cite{Padmanabhan:1987dg} and even crosses the phantom barrier at some point \cite{Cataldo:2005qh,Brevik:2006wa,Cruz2017,NCruz2017}. Possible quantum effects for those viscous phantom models have been also analysed close to future singularities \cite{Brevik:2008xv}.  Moreover, such effects may play an important role in the evolution of relativistic fluids, see for example Ref.~\cite{Nunes}, where was shown that a small amount of viscosity in cold dark matter fluid seems to alleviate a couple of discrepancies in the values obtained for some cosmological parameters when large scale structure (LSS) and Planck CMB data are used. Note also that baryons play an essential role in the formation of structure and can present a friction much larger than the possible coming from dark matter, as observed in weak lensing of cluster mergers (see Ref.~\cite{Clowe:2003tk}). Nevertheless, the density of baryons at current times is small enough to consider its possible dissipative terms negligible in comparison to the dark matter viscosity, since the bulk viscosity is usually assumed to be proportional to the energy density.  \\

To perform this type of analysis, the Eckart approach provides an straightforward way to analyse such properties in cosmology, as bulk viscosity is introduced through an effective contribution to the pressure \cite{Cataldo:2005qh}. As the total effective pressure determines the behaviour of the universe expansion, in the presence of bulk viscosity effective pressure may become negative even in the case of a fluid fulfilling the energy conditions, providing a physically acceptable way to explain the late-time acceleration. In this work, we analyse the effects of bulk viscosity following the Israel-Stewart theory, which establishes the formalism for relativistic non-perfect fluids \cite{Israel1979}, which determines the dynamics of the viscous pressure. It is important to mention that the condition the near equilibrium condition is demanded in the thermodynamics approaches of relativistic viscous fluids, which means that the viscous stress must be lower than the equilibrium pressure of the fluid. Nevertheless, if we require accelerated expansion only due to the negativeness of the bulk viscosity effective pressure, then the near equilibrium condition is not fulfilled. We shall discuss below that the introduction of a cosmological constant can allow to satisfy the above condition. The price to pay is, of course, to abandon unified dark matter models with dissipation as a consistent models to describe the evolution of the universe. Hence, here we are assuming the presence of a cosmological constant  while dark matter is described by a fluid with dissipative effects according to the Israel-Stewart theory. The existence of de Sitter solutions is analysed and discussed depending on the form of the dissipative terms. Then, by using Supernovae Ia data, the free parameters of the theory are constrained and compared to $\Lambda$CDM model.\\

The paper is organised as follows: in section \ref{sectIS}, the Israel-Stewart formalism is introduced. Section \ref{sectDS} is devoted to the analysis on the existence of de Sitter solutions. In section \ref{sectNE}, some discussion is provided about the condition to be near the equilibrium. Section \ref{sect_sneia} provides the fits of the model by using Sne Ia data, while Section \ref{sect_results} discuss its results. Finally, section \ref{sect_conclusions} focus on the conclusions of the paper.


\section{Israel-Stewart formalism with cosmological constant}
\label{sectIS}

Let us introduce the formalism we follow along the paper. As usual, we focus on a flat Friedmann-Lema\^itre-Robertson-Walker universe, whose metric can be expressed as
\be
ds^2=-dt^2+a^2(t)\sum^3_{i=1} dx_i^2\ ,
\label{FLRWm}
\ee
where $a(t)$ is the scale factor and $t$ is the cosmic time. In what follows we assume that the universe contains dark energy in the form of a positive cosmological constant and dark matter, expressed as a perfect fluid with dissipative effects during cosmic evolution. Hence,  we assume a barotropic equation of state (EoS) for dark matter, $p=\omega \rho $, where $p$ is the barotropic pressure and $0\leq\omega <1$. For a flat FLRW universe, the constraint equation is given by
\begin{eqnarray}
H^{2}=\frac{\dot{a}^2}{a^2}=\frac{\kappa^2}{3}\rho+\frac{\Lambda}{3}\ , \label{constraint}
\end{eqnarray}
while the continuity equation for the dissipative fluid is defined as
\begin{eqnarray}
\dot{\rho}+3H\left[\left(1+\omega\right)\rho+\Pi\right]=0. \label{conservation}
\end{eqnarray}
In the Israel-Stewart framework, the transportation equation for the viscous pressure $\Pi $ is given by \cite{Israel1979}
\begin{eqnarray}
\tau\dot{\Pi}+\left(1+\frac{1}{2}\tau\Delta\right)\Pi=-3\zeta\left(\rho\right)H, \label{eqforPi}
\end{eqnarray}
where dots denote derivatives with respect to the cosmic time, $\tau$ is the relaxation time, $\zeta(\rho)$ is the bulk viscosity coefficient, for which we assume the usual dependence upon the energy density $\rho$, $H$ is the Hubble parameter, and $\Delta$ is defined by the expression
\begin{eqnarray}
\Delta=3H+\frac{\dot{\tau}}{\tau }-\frac{\dot{\zeta}}{\zeta}-\frac{\dot{T}}{T}, \label{Delta}
\end{eqnarray}
where $T$ is the barotropic temperature, which takes the form $T=T_{0}\rho^{\omega/\left(\omega+1\right)}$ (Gibbs integrability condition when $p=\omega\rho$) with $T_{0}$ being a positive parameter. Note that for $\omega\sim 0$, the temperature turns out $T=T_0$, which can be set to accomplish the requirements from the different dark matter candidates suggested in the literature (for a review see \cite{Feng:2010gw}). The dark matter EoS, $\zeta(\rho)$ and the relaxation time, $\tau$, are related by ~\cite{Maartens:1996vi}
\begin{eqnarray}
\frac{\zeta}{\left(\rho+p\right)\tau}=c_{b}^{2}, \label{relaxationtime}
\end{eqnarray}
where $c_{b}$ is the speed of bulk viscous perturbations (non-adiabatic contribution to the speed of sound in a dissipative fluid without heat flux or shear viscosity), since the dissipative speed of sound, $V$ is given by $V^{2}= c_{s}^{2}+c_{b}^{2}$, where $c_{s}^{2}=(\partial p/\partial \rho)_{s}$ is the adiabatic contribution. For a barotropic fluid $c_{s}^{2}=\omega$ and the speed of sound for the viscous perturbations can be expressed in terms of the barotropic fluid as $c_{b}^{2}=\epsilon\left(1-\omega\right)$ with $0<\epsilon\leq 1$. From here on, since the second law of thermodynamics should be satisfied, we assume $\zeta=\zeta_{0}\rho^{s}$ as positive, such that  $\zeta _{0}$ should be a positive constant ~\cite{Weinberg1971}. Following the Eckart formalism for instance, $s$ is kept as an arbitrary parameter. Then, from (\ref{relaxationtime}) the relaxation time results to be
\begin{eqnarray}
\tau=\frac{1}{\epsilon\left(1-\omega^{2}\right)}\frac{\zeta}{\rho}=\frac{\zeta_{0}}{\epsilon\left(1-\omega^{2}\right)}\rho^{s-1}. \label{relaxationtime1}
\end{eqnarray}
In order to find a differential equation in terms solely of the Hubble parameter, $H$, we evaluate the expressions $\dot{\tau}/\tau, \,\, \dot{\zeta}/\zeta$ and $\dot{T}/T$ from (\ref{Delta}). By using (\ref{constraint}), we obtain the following expressions 
\begin{eqnarray}
\frac{\dot{\tau}}{\tau}=\frac{6\left(s-1\right)}{3H^{2}-\Lambda}H\dot{H}, \nn
\frac{\dot{\zeta}}{\zeta}=\frac{6s}{3H^{2}-\Lambda}H\dot{H}, \nn
\frac{\dot{T}}{T}=\frac{6\omega}{\left(\omega+1\right)\left(3H^{2}-\Lambda \right)}H\dot{H}.  \label{TpuntoT}
\end{eqnarray}
By substituting the expressions (\ref{TpuntoT}) in equation (\ref{Delta}), the expression for $\Delta$ is rewritten as
\begin{eqnarray}
\Delta=\Delta\left(H\right)=\frac{3H}{\delta\left(\omega\right)}\left(\delta\left(\omega\right)
-\frac{\dot{H}}{H^{2}-\Lambda/3}\right), \label{Deltanew}
\end{eqnarray}
where
\begin{eqnarray}
\delta\left(\omega\right)\equiv\frac{3}{4}\left(\frac{\omega+1}{\omega+1/2}\right). \label{delta}
\end{eqnarray}
From (\ref{constraint}) and (\ref{conservation}), we can obtain the following expression for the viscous pressure
\begin{eqnarray}
\Pi=-\left[2\dot{H}+\left(1+\omega\right)\left(3H^{2}-\Lambda\right)\right], \label{Pi}
\end{eqnarray}
Finally, by using the equation (\ref{eqforPi}), the following differential equation for $H$ is obtained
\begin{widetext}
\begin{eqnarray}
\frac{\zeta_{0}}{\epsilon\left(1-\omega^{2}\right)}\left(3H^{2}-\Lambda\right)^{s-1}\left[2\ddot{H}+6\left(1+\omega\right)H\dot{H}\right]-3\zeta_{0}\left(3H^{2}-\Lambda\right)^{s}H \nonumber \\
+\left[1+\frac{\zeta _{0}}{2\epsilon\left(1-\omega^{2}\right)}\left(3H^{2}-\Lambda\right)^{s-1}\Delta \left( H\right)\right]\left[2\dot{H}+\left(1+\omega\right)\left(3H^{2}-\Lambda\right)\right]=0,
\label{eqforH}
\end{eqnarray}
\end{widetext}
where recall that $\Delta \left( H\right) $ is defined in (\ref{Deltanew}). As discussed in Ref.~\cite{Cruz2017}, in absence of a cosmological constant, $\Lambda =0$, and for the special case $s=1/2$, (\ref{eqforH}) has a phantom solution of the form $H\left(t\right)=A\left(t_{s}-t\right)^{-1}$, with $A>0$ and the restriction $ 0<\omega <1/2$. 

\section{De Sitter solutions}
\label{sectDS}

Here we analyse the existence of de Sitter solutions of the equation (\ref{eqforH}). Then, we assume a constant Hubble parameter $H=H_{0}$, such that equation (\ref{eqforH}) becomes a simple algebraic equation for $H_{0}$, leading to
\begin{eqnarray}
H_{0}\left(3H_{0}^{2}-\Lambda\right)^{s-1}=\frac{A}{\zeta_{0}}, \label{eqforHconst}
\end{eqnarray}
where for simplicity we have introduced the definition
\begin{eqnarray}
A\equiv\frac{1}{3}\left[\frac{2\epsilon\left(1-\omega^{2}\right)}{2\epsilon\left(1-\omega\right)-1}\right], \label{defofA}
\end{eqnarray}
In absence of matter, the equation leads to the two well known solutions,
\begin{eqnarray}
H_{0}=\pm\sqrt{\frac{\Lambda}{3}}, \label{deSitter}
\end{eqnarray}
i.e., are the standard de Sitter solutions. However, let us investigate the existence of other de Sitter solutions in the presence of the dissipative dark matter fluid. To do so, we solve equation (\ref{eqforHconst}) for some values of the parameter $s$. Nevertheless, firstly we explore the conditions for which $A>0$, $A<0$ and the consequences of $A=0$. Note that the parameter $\zeta_{0}$ does not plays any role in this analysis because is always a positive constant.  

If $A=0$, then equation (\ref{eqforHconst}) has the solutions by (\ref{deSitter}), independently of the value of $s$, and the trivial solution $H_{0}=0$. By discarding the possibility of $A=0$, we have $\epsilon\neq 0$ and $\omega\neq 1$, which recall that $0<\epsilon\leq 1$ and $0\leq\omega<1$. In ~\cite{NCruz2017}, they discuss the possibility of $\omega=1$ in the context of the Israel-Stewart theory without a cosmological constant.

If $A>0$, then from (\ref{defofA}), we see that $2\epsilon\left(1-\omega \right)-1>0$ or $\epsilon>1/2\left(1-\omega\right)$, since $\left(1-\omega^{2}\right)>0$, $\left(1-\omega\right)>0$ and $\epsilon>0$. Besides $\epsilon \leq1$,  we need to impose $1/2\left(1-\omega\right)<1$ or $\omega<1/2$, which can be summarised in the following constraints
\begin{eqnarray}
\frac{1}{2}\leq\frac{1}{2\left(1-\omega\right)}<\epsilon\leq 1 \;\; \textup{with} \;\; 0\leq\omega<\frac{1}{2}. \label{constraintApositiv}
\end{eqnarray}

If $A<0$,  (\ref{defofA}) gives the following constraints,
\begin{eqnarray}
0<\epsilon<\frac{1}{2\left(1-\omega\right)} \;\; \textup{with} \;\; 0\leq\omega<\frac{1}{2}, \nonumber \\ 
\textup{or}, \;\: 0<\epsilon\leq 1 \;\; \textup{with} \;\; \frac{1}{2}\leq\omega<1. \label{constraintAnegativ}
\end{eqnarray}

Hence, the possible de Sitter solutions can be explored now in terms of the value of $s$.\\

\textbf{i) $s=1$:} For this value of $s$, it is straightforward to obtain the solution
\begin{eqnarray}
H_{0}=\frac{A}{\zeta_{0}}. \label{solforsigual1}
\end{eqnarray}
An expanding solution with $H_{0}>0$ is obtained if $A>0$, so in this case we have the constraints indicated in the (\ref{constraintApositiv}).\\

\textbf{ii) $s=1/2$:} In this case the solution of (\ref{eqforHconst}) has the form
\begin{eqnarray}
H_{0}=\sqrt{\Lambda\left(3-\frac{\zeta_{0}^{2}}{A^{2}}\right)^{-1}}, \label{solforsigual1/2}
\end{eqnarray}
where we haave chosen the positive sign that represents an expanding solution. For a positive cosmological constant, the real solution is obtained if $3-\zeta_{0}^{2}/A^{2}>0$ and leads to
\begin{eqnarray}
\left(\zeta_{0}+\sqrt{3}A\right)\left(\zeta_{0}-\sqrt{3}A\right)<0, \label{inequationforxi}
\end{eqnarray}
thus, the interval of $\zeta_{0}$ is given by
\begin{eqnarray}
-\sqrt{3}A<\zeta_{0}<\sqrt{3}A. \label{constraintxi}
\end{eqnarray}
Note that this result is invariant under the choice of $A>0$ or $A<0$, so it is not possible in principle to add the constraints (\ref{constraintApositiv}) or (\ref{constraintAnegativ}). To avoid problems of interpretation and in order to consider the two possibilities, we take the absolute value of $A$ and assume that $\zeta_{0}>0$, such that the constraint (\ref{constraintxi}) is rewritten as
\begin{eqnarray}
0<\zeta_{0}<\sqrt{3}|A|. \label{constraintxiabs}
\end{eqnarray}
\\
\textbf{iii) $s=0$:} In this case the solution of (\ref{eqforHconst}) has the form
\begin{eqnarray}
H_{0}=\frac{\zeta_{0}}{6A}\left(1\pm\sqrt{1+\frac{12A^{2}\Lambda}{\zeta_{0}^{2}}}\right). \label{solforsigual0}
\end{eqnarray}
For a positive cosmological constant, the argument of the root in (\ref{solforsigual0}) is always positive, then this solution is a real number. For the positive sign in (\ref{solforsigual0}), since $\sqrt{1+12A^{2}\Lambda/\zeta_{0}^{2}}>1$, the expanding solution is obtained whether $A>0$, which consequently imposes the condition (\ref{constraintApositiv}). On the other hand, for the negative sign, the expanding solution is obtained whether $A<0$, what leads to the constraint (\ref{constraintAnegativ}).\\

\textbf{iv) $s=-1/2$:} In this case we can rewrite equation (\ref{eqforHconst}) as
\begin{eqnarray}
x^{3}-\frac{\zeta_{0}^{2}}{3A^{2}}x-\frac{\Lambda\zeta_{0}^{2}}{3A^{2}}=0, \label{solforsigual-1/2}
\end{eqnarray}
where we have defined
\begin{eqnarray}
x=3H_{0}^{2}-\Lambda. \label{defofx}
\end{eqnarray}
Equation (\ref{solforsigual-1/2}) can be solved by using the Cardano's method. In this method we make a change of variables, $x=u+v$ which reduces the above cubic equation to the following quadratic equation
\begin{eqnarray}
z^{2}-\frac{\Lambda\zeta_{0}^{2}}{3A^{2}}z+\frac{\zeta_{0}^{6}}{729A^{6}}=0, \label{eqforz}
\end{eqnarray}
where $z=v^{3}$, $uv=\frac{\zeta_{0}^{2}}{9A^{2}}$ and $u^{3}+v^{3}=\frac{\Lambda\zeta_{0}^{2}}{3A^{2}}$ and from (\ref{eqforz}) we obtain
\begin{eqnarray}
u=\left(\frac{\Lambda\zeta_{0}^{2}}{6A^{2}}\right)^{1/3}\left(1+\sqrt{\Delta}\right)^{1/3}, \nn
v=\left(\frac{\Lambda\zeta_{0}^{2}}{6A^{2}}\right)^{1/3}\left(1-\sqrt{\Delta}\right)^{1/3}\ , \label{defofv}
\end{eqnarray}
where $\Delta$ is given by
\begin{eqnarray}
\Delta=1-\frac{4\zeta_{0}^{2}}{81\Lambda^{2}A^{2}}. \label{defofdelta}
\end{eqnarray}
 Solutions for (\ref{solforsigual-1/2}) depends on the sign of $\Delta$. If $\Delta>0$ then (\ref{solforsigual-1/2}) has one real and two complex solution, if $\Delta=0$ it has three real solutions where two of them are equal and if $\Delta<0$ it has three different real solutions. In the case $\Delta>0$, equation (\ref{defofdelta}) leads to
\begin{eqnarray}
\left(\Lambda+\frac{2\zeta_{0}}{9A}\right)\left(\Lambda-\frac{2\zeta_{0}}{9A}\right)>0, \label{inequationforlambda}
\end{eqnarray}
and $\Lambda$ is constrained to be
\begin{eqnarray}
\Lambda<-\frac{2\zeta_{0}}{9A} \;\; \textup{or} \;\; \Lambda>\frac{2\zeta_{0}}{9A}.
\label{constraintlambda}
\end{eqnarray}
Note that as in the case $s=1/2$, this result is independent of the sign of $A$. So, by assuming the absolute value of $A$ and considering $\Lambda>0$, we can rewrite (\ref{constraintlambda}) in the form
\begin{eqnarray}
\Lambda>\frac{2\zeta_{0}}{9|A|}.
\label{constraintlambdaabs}
\end{eqnarray}
The only real solution of (\ref{solforsigual-1/2}) is given by $x=u+v$ where it is clear that $x>0$. From Eqs.~(\ref{defofx}) and (\ref{defofv}), we obtain a Hubble parameter given by
\begin{eqnarray}
H_{0}=\sqrt{\frac{1}{3}(u+v)+\frac{\Lambda}{3}}, \label{Hdeltapositiv}
\end{eqnarray}
where we have chosen the positive sign for representing the expanding solution. No more restriction are required because the argument of the root in the above equation is positive.

In the case $\Delta=0$, equation (\ref{defofdelta}) leads to
\begin{eqnarray}
\Lambda=\pm\frac{2\zeta_{0}}{9A}\ . \label{constrainlambdacero}
\end{eqnarray}
And the positive sign and a positive cosmological constant requires $A>0$, leading to (\ref{constraintApositiv}). On the other hand, the negative sign and a positive cosmological constant requires $A<0$, which is provided as far as (\ref{constraintAnegativ}) is satisfied. A real solution for (\ref{solforsigual-1/2}) is given by $x=u+v$, so from (\ref{defofx}) we have the following Hubble parameter
\begin{eqnarray}
H_{0}=\sqrt{\pm\frac{8\zeta_{0}}{27A}}, \label{Hdeltacero}
\end{eqnarray}
which is a real number as $\Lambda>0$. The other two solutions of (\ref{solforsigual-1/2}) have the form $x_{1}=x_{2}=\mp\frac{\zeta_{0}}{3A}$ but from (\ref{defofx}), they lead to a complex Hubble parameter.

In the case $\Delta<0$,  we see that $\Lambda$ has to satisfy the range (\ref{constraintlambdaabs}) and $\Lambda>0$, we can write
\begin{eqnarray}
0<\Lambda<\frac{2\zeta_{0}}{9|A|}.
\label{constrainlambdaabsneg}
\end{eqnarray}
The solutions of (\ref{solforsigual-1/2}) can be written in the form
\begin{eqnarray}
x_{k}=\frac{2\zeta_{0}}{3A}\cos{\left[\frac{1}{3}\arccos{\left(\frac{9A}{2\zeta_{0}}\Lambda\right)}+\frac{2k\pi}{3}\right]}, \label{xdeltanegativ}
\end{eqnarray}
where $k\in\{0,1,2\}$. The Hubble parameter takes the form
\begin{eqnarray}
H_{0,k}=\sqrt{\frac{2\zeta_{0}}{9A}\cos{\left[\frac{1}{3}\arccos{\left(\frac{9A}{2\zeta_{0}}\Lambda\right)}+\frac{2k\pi}{3}\right]+\frac{\Lambda}{3}}}, \label{Hdeltanegativ}
\end{eqnarray}
where again $k\in\{0,1,2\}$. We need to determine what solutions are real. First, note that the argument of the $\arccos$ function in (\ref{Hdeltanegativ}) is always real because we have to fulfill the condition (\ref{constrainlambdaabsneg}) and this $\arccos$ function is bounded by $0$ and $\pi$, thus
\begin{eqnarray}
\frac{2k\pi}{3}\leq\frac{1}{3}\arccos{\left(\frac{9A}{2\zeta_{0}}\Lambda\right)}+\frac{2k\pi}{3}<\frac{\pi}{3}+\frac{2k\pi}{3}. \label{arccosbounded}
\end{eqnarray}
From the above equation, if $k=0$, then the $\cos$ function will remain in the first quadrant where is always positive and the solution for $H$ will be real for $A>0$. If $k=1$, the $\cos$ function will remains in the second quadrant where is always negative and the solution for $H$ will be real for $A<0$. Finally if $k=2$ the $\cos$ function will remain in the third and fourth quadrant, where is negative and positive respectively. In the third quadrant clearly we have (note that the function is displaced by $4\pi/3$)
\begin{eqnarray}
0\leq\frac{9A}{2\zeta_{0}}\Lambda\leq 1, \label{thirdquadrant}
\label{131}
\end{eqnarray}
i.e., in this area we have $A>0$ but a real solution requires $A<0$, which it is not possible. On the other hand, for the fourth quadrant we have
\begin{eqnarray}
-1\leq\frac{9A}{2\zeta_{0}}\Lambda\leq 0, \label{fourthquadrant}
\end{eqnarray}
which consequently leads to $A<0$ but a real solution requires $A>0$. So the only real Hubble parameter is given by (\ref{Hdeltanegativ}) with $k=0,1$.

By observing these particular solutions, where de
Sitter behaviour is assumed for all times, we can learn some general issues about
the values of the parameters involved in the model.  We shall
restrict to $\omega\approx 0$ since our main interest is the
behavior of a cold dark matter, or some sort of warm dark matter,
with dissipation and the comparison with the $\Lambda CDM$ model. In
the case $s=1$ the constraint for $\epsilon$ is $1/2 < \epsilon < 1$
is provided by Eq. (\ref{constraintApositiv}) and the Hubble parameter is given by Eq(\ref{solforsigual1}).
Therefore, we have an scenario with a great non-adiabatic
contribution to the speed of sound and, on the other hand, a Hubble
parameter which decreases as $\xi_{0}$ and does not depend on the
cosmological constant. Therefore, this solution have two non
desirable physical behaviour.

In the case of $s=1/2$, there is no constraint on $\epsilon$ and for
$\omega\approx 0$, and we obtain from Eq(\ref{defofA}) that $\xi_{0}< 2\epsilon/
\sqrt{3}(2\epsilon -1)$. For the Hubble parameter the
expression given in Eq.~(\ref{solforsigual1/2}) reduces to the usual form $H_{0}=
\sqrt{\Lambda/3}$ (de Sitter solution) when $\xi_{0}$ goes to zero.
Then, this solution can represent well the asymptotic desirable
behaviour of a general solution for the Eq.~(\ref{eqforH}).

The case of $s=0$ corresponds to a constant bulk viscosity. The
expanding solution has also the constraint $1/2 < \epsilon < 1$, and
the arguments mentioned above applied. The Hubble parameter, which
goes to zero when $\xi_{0}\rightarrow 0$, also does not reduce to the
asymptotical behaviour of $\Lambda CDM$.

When $s=-1/2$ the solution leads to a lower bound for the
cosmological constant given by Eq.~(). Then a positive cosmological
constant and the expression for $|A|$ with $\omega=0$, leads to the
constraint $\epsilon >1/2$.  The Hubble parameter, given by Eq.~(\ref{constraintlambdaabs})
or Eq.~(\ref{Hdeltanegativ}), for the special case with $\Delta =0$, goes to a de
Sitter solution when $\xi_{0}$ goes to zero. Hence, the weakness of
this solution is also the requirement of great non-adiabatic
contribution to the speed of sound.

In summary, the study of de Sitter type solutions for some simple
values of the parameter $s$ seems to indicate that the case of
$s=1/2$, which has the advantages of simplifying the Eq.~(\ref{eqforH}), might be
investigated further in order to find an exact solution. In section V
we investigate our model in terms of the constraints imposed by the
astronomical data, particularly by Supernovae Ia. The parameters $s,\xi_{0}$ and $\Lambda$ will set
free, which implies to face more general solutions of the Israel-Stewart theory for arbitrary
parameter $s$, which requires numerical resources.
\section{Near equilibrium condition}
\label{sectNE}

As previously discussed in \cite{Maartens1995}, within the context of dissipative inflation, the condition to have an accelerating expansion imposes negativeness on the viscous pressure $\Pi$. Let us consider the following equation on the second derivative of the scale factor
\begin{eqnarray}
\frac{\ddot{a}}{a}=-\frac{1}{6}\left(\rho+3P_{eff}\right)+\frac{\Lambda}{3}, \label{rr}
\end{eqnarray}
where $P_{eff}=p+\Pi$. By imposing $\ddot{a}>0$ in (\ref{rr}) and taking $\Lambda =0$, it yields
\begin{eqnarray} 
-\Pi >p+\frac{\rho}{3}. \label{Pieqa}
\end{eqnarray}
So the inequality (\ref{Pieqa}) implies that the viscous stress is greater than  the equilibrium pressure $p$ of the fluid. The non-causal approach of Eckart and the causal of Israel-Stewart assume a near equilibrium regime that must fulfill the condition
\begin{eqnarray}
\left|\frac{\Pi}{p}\right|\ll 1, \label{nearequilibrium}
\end{eqnarray}
but in order to obtain an accelerating expansion, the fluid has to be far from equilibrium. This situation changes if the cosmological constant is included, in whose case the condition $\ddot{a}>0$ leads to
\begin{eqnarray}
-\Pi >\frac{-2\Lambda}{3}+p +\frac{\rho}{3}, \label{Pieqa1}
\end{eqnarray}
so the viscous stress is not necessarily greater than the equilibrium pressure $p$ in order to have an accelerating expansion and near equilibrium condition may be fulfilled in some cases. It is important to note that to fulfil the equilibrium condition, a positive cosmological constant is required.

\section{Fitting the Israel-Stewart model to Supernova Ia data} \label{sect_sneia}

Let us now compare the previous model with observational data. To do so, here we use the Union 2.1 SN catalogue  (see \cite{Suzuki:2011hu}), which contains $N_{\text{SN}}=557$ type Ia supernovas with redshifts $0.015\le z\le 1.4.$ The catalogue provides the corresponding redshift and the values of the distance modulus for each SN as well as the corresponding errors $\sigma_{\mu_{\text{obs}}(z)}.$ 

Firstly, we define the corresponding magnitudes involved in the analysis. The free luminosity distance is given by
\begin{equation}
D_L(z;\Omega_m,\alpha_i)= (1+z) \int_0^z {\rm d}z'\frac{H_0}{H(z';\Omega_m,\alpha_i)}\ .
\label{SN1} 
\end{equation}
Here $\Omega_m$ is the corresponding matter density and $\alpha_i$ are the free parameters of the model. Then, the distance modulus, used to fit to the data, is defined as follows
\begin{equation}
\mu_{\text{theo}}(z;\Omega_m,\alpha_i)=\bar\mu+5\log_{10} \left[D_L(z;\Omega_m,\alpha_i)\right]\ ,
\label{SN2} 
\end{equation}
where $\bar \mu)=-5\log_{10}\left[{H_0\over c}\right]+25$ is a nuisance parameter. Here we use assume a Gaussian distribution,
\begin{equation}
\mathcal{L}= {\cal N} {\rm e}^{- \chi^2/2}\ , 
\label{SN3} 
\end{equation} 
where   ${\cal N}$ is a normalisation factor and the function $\chi^2$ is defined as follows
\begin{eqnarray}
\chi^2 =\sum_{i=1}^{N} \frac{(\mu_{\text{obs}}(z_i) - \mu_{\text{theo}}(z_i;{\bar \mu}, \Omega_m, \alpha_i))^2} {\sigma_{\text{obs}}^2(z_i)}\,
\label{SN4} 
\end{eqnarray}
To find the best fits for the free parameters, we use the technique of the minimum $\chi^2_{\text{min}}$ which consequently maximise the probability distribution (\ref{SN3}). In order to reduce the number of the free parameters, and particularly in order to marginalise over the nuisance parameter $\bar\mu$, the function (\ref{SN4}) can be expanded as \cite{Leanizbarrutia:2014xta,Lazkoz:2005sp}
\begin{equation}
\chi^2 (\Omega_m, \alpha)= A - 2{\bar \mu}B  + {\bar \mu}^2C\ ,
\label{SN5} 
\end{equation}
where
\begin{eqnarray}
A(\Omega_m, \alpha_i)&=&\sum_{i=1}^{N_{\text{SN}}} \frac{(\mu_{\text{obs}}(z_i) - \mu_{\text{theo}}(z_i ;{\bar \mu}=0, \Omega_m, \alpha_i))^2}{\sigma_{\mu_{\text{obs}} (z_i)}^2} 
\label{SN6.1} \nonumber \\
B(\Omega_m, \alpha_i)&=&\sum_{i=1}^{N_{\text{SN}}}\frac{(\mu_{\text{obs}}(z_i) - \mu_{\text{theo}}(z_i ;{\bar \mu}=0, \Omega_m, \alpha_i))}{\sigma_{\mu_{\text{obs}}(z_i)}^2} 
\label{SN6.2} \nonumber \\
C&=&\sum_{i=1}^{N_{\text{SN}}}\frac{1}{\sigma_{\mu_{\text{obs}}(z_i)}^2 } 
\label{SN6.3}
\end{eqnarray} 
Hence, by minimising the expression (\ref{SN5}) with respect to $\bar\mu$, one obtains $\bar\mu=B/C$ and the $\chi^2$ finally reduces to
\begin{equation}
{\tilde\chi}^2(\Omega_m, \alpha_i)=A(\Omega_m, \alpha_i)- \frac{B^2(\Omega_m, \alpha_i)}{C}\,. 
\label{SN7}
\end{equation}
This is the expression for the $\chi^2$ that we will use in our MCMC analyses. For simplicity, we omit the tilde from now on. In order to calculate the distance modulus, equation (\ref{eqforH}) has to be solved for the Hubble parameter. To simplify the calculations and compute the distance modulusin a better way, we consider the redshift $z$ as the independent variable instead of the cosmic time $t$
\be
1+z=\frac{1}{a}\ .
\label{SN8}
\ee
Here $a$ is the scale factor and we have assumed $a_0=1$ as the reference value of the scale factor evaluated today. Then, equation (\ref{eqforH}) turns out
\begin{widetext}
\begin{eqnarray}
&&\frac{\zeta_{0}}{\epsilon\left(1-\omega^{2}\right)}\left(3H^{2}-\Lambda\right)^{s-1}(1+z)HH^{\prime}\left[2\left(H+(1+z)H^{\prime}+(1+z)H\frac{H^{\prime\prime}}{H^{\prime}}\right)-6\left(1+\omega\right)H\right]-3\zeta_{0}\left(3H^{2}-\Lambda\right)^{s}H \nonumber \\
&+&\left[1+\frac{\zeta _{0}}{2\epsilon\left(1-\omega^{2}\right)}\left(3H^{2}-\Lambda\right)^{s-1}\Delta \left( H\right)\right]\left[-2(1+z)HH^{\prime}+\left(1+\omega\right)\left(3H^{2}-\Lambda\right)\right]=0,
\label{SN9}
\end{eqnarray}
\end{widetext}
The set of free parameters are $\zeta_0$, $s$ and $\Omega_m$, since we are considering pressureless fluid, such that we are assuming that our model is well described by an equation of state parameter $w$ closely to zero, so negligible. Moreover, the parameter $\epsilon$, which defines the speed of sound of the viscous perturbations, can take any value in the range $0<\epsilon\leq 1$, where $\epsilon=1$ would correspond to the speed of light. As shown in the section below, we analyse the model for several values of $\epsilon$, which affect the free parameters and the corresponding errors. To implement the MCMC, we use the Metropolis--Hasting algorithm by running several chains and analysing the convergence of them. 

\begin{figure*} 
\begin{center}
\includegraphics[width=0.29\textwidth]{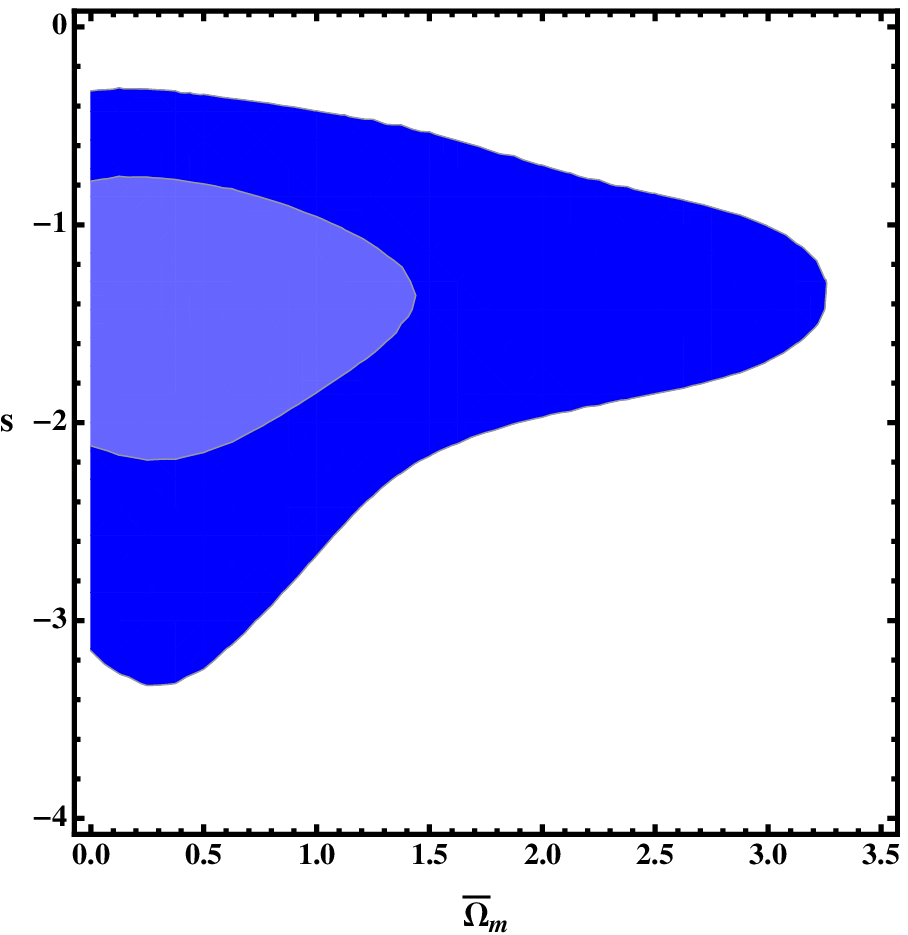}
\includegraphics[width=0.3\textwidth]{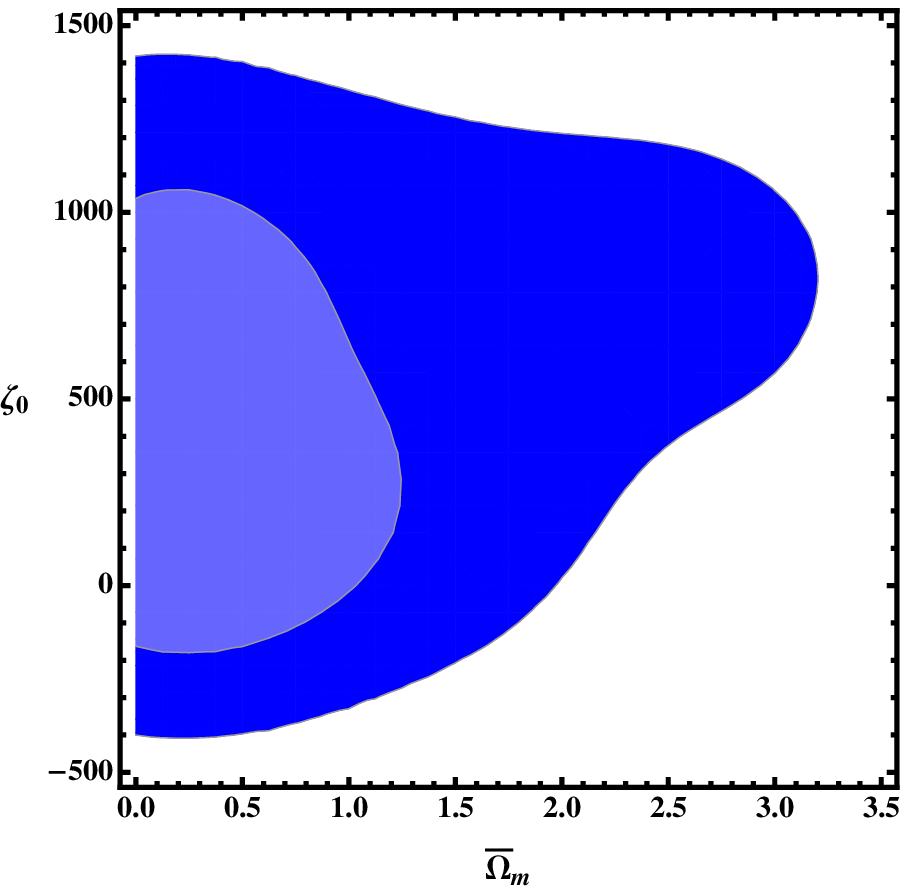}
\includegraphics[width=0.3\textwidth]{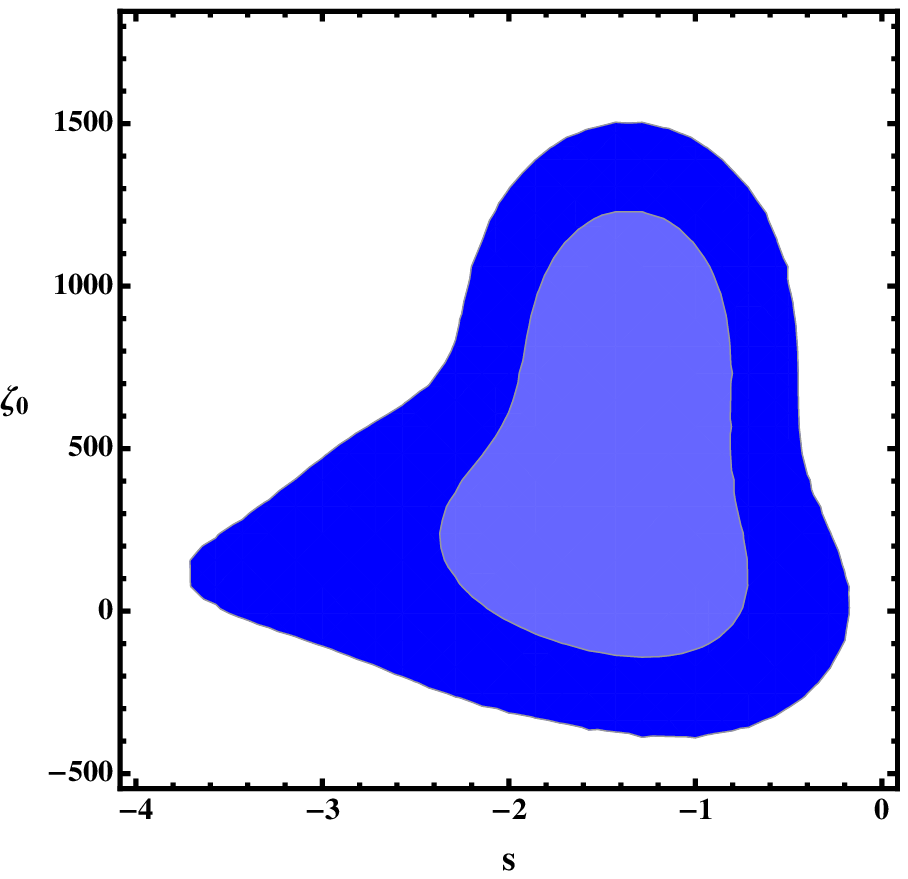}
\end{center}
\caption{Contour plots for the free parameters $\{\zeta_0, \Omega_m, s\}$ when fitting the Israel-Stewart model with data from Union 2.1. This case representes $\epsilon=1$.}
\label{Fig1} 
\end{figure*}

\begin{figure*} 
\begin{center}
\includegraphics[width=0.29\textwidth]{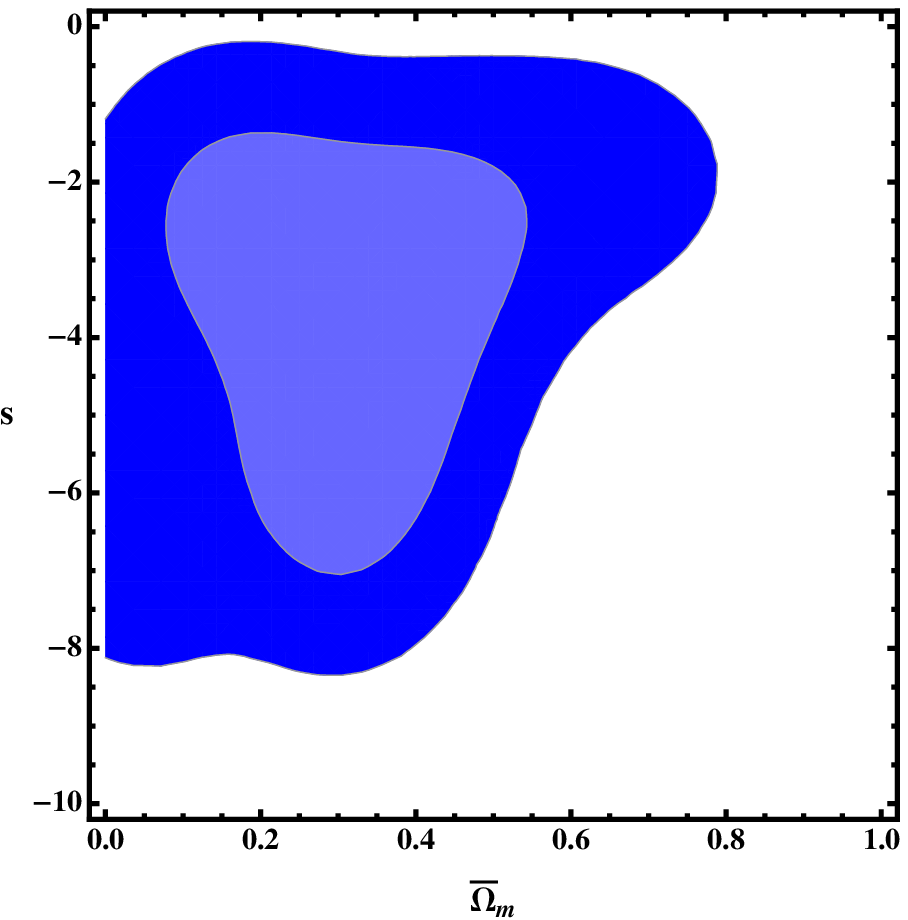}
\includegraphics[width=0.3\textwidth]{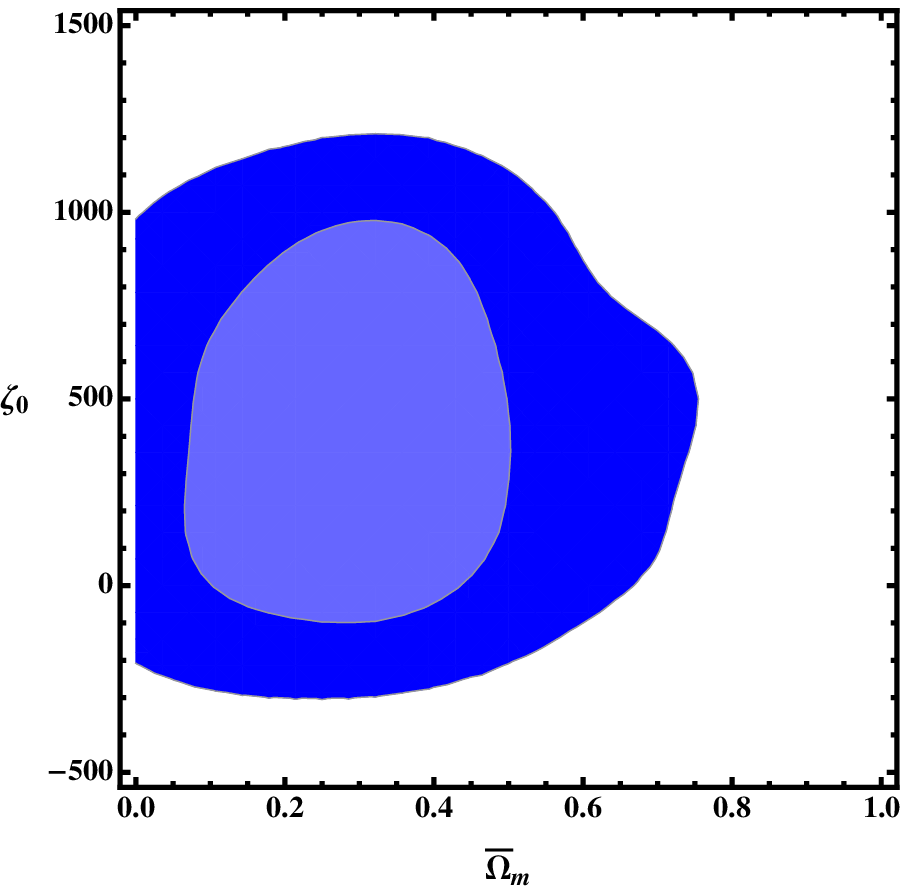}
\includegraphics[width=0.3\textwidth]{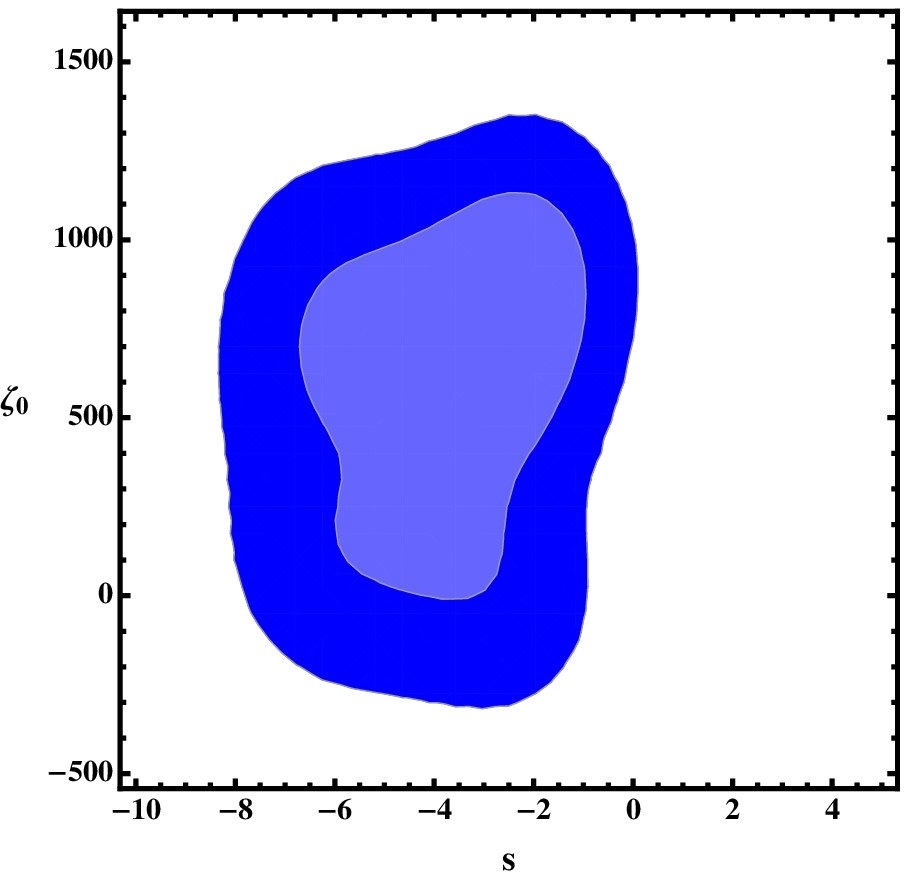}
\end{center}
\caption{Contour plots for the free parameters $\{\zeta_0, \Omega_m, s\}$ when fitting the Israel-Stewart model with data from Union 2.1. This case representes $\epsilon=0.5$.}
\label{Fig1a} 
\end{figure*}

\begin{figure*} 
\begin{center}
\includegraphics[width=0.29\textwidth]{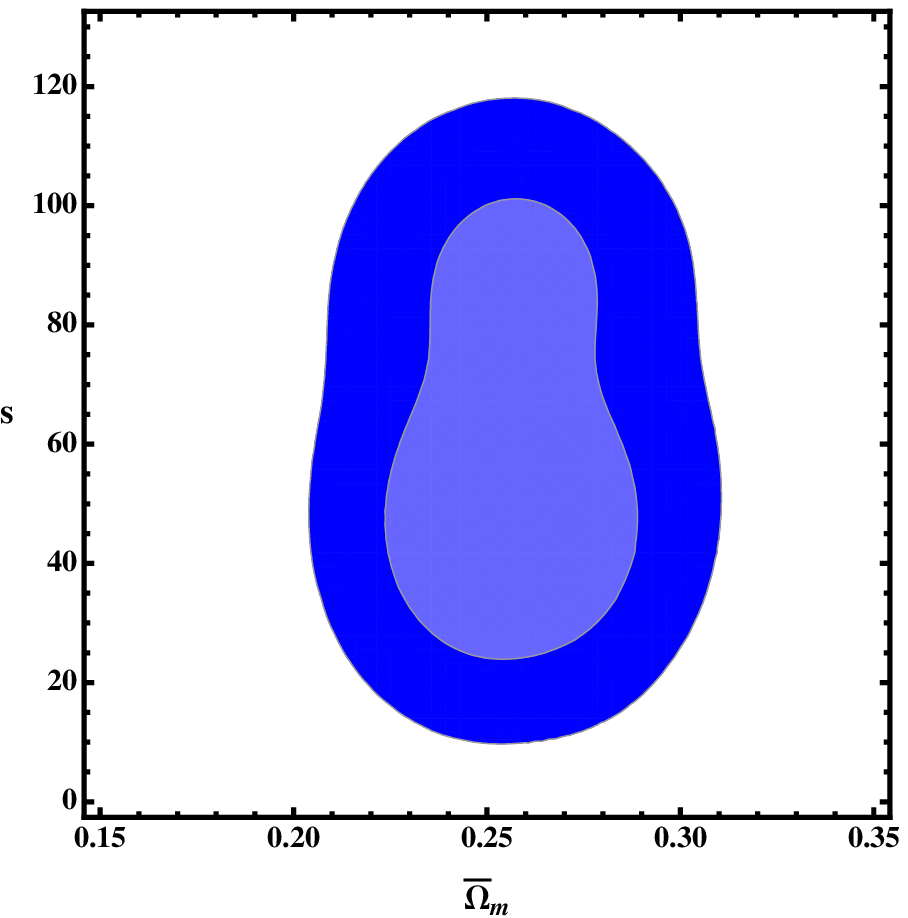}
\includegraphics[width=0.3\textwidth]{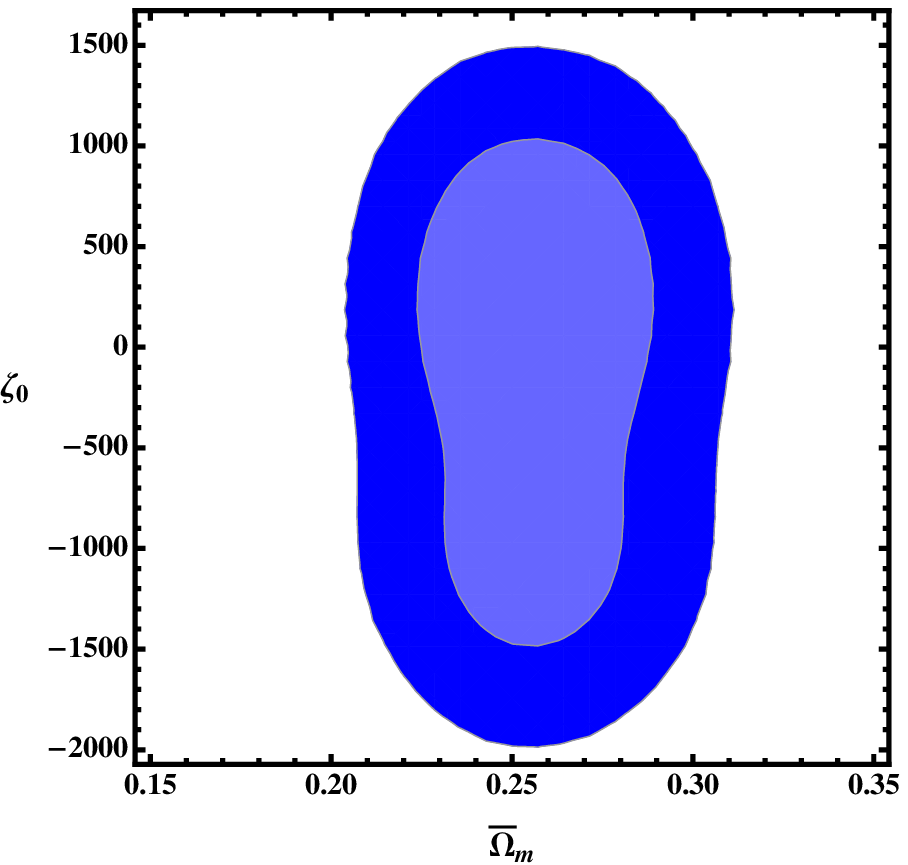}
\includegraphics[width=0.3\textwidth]{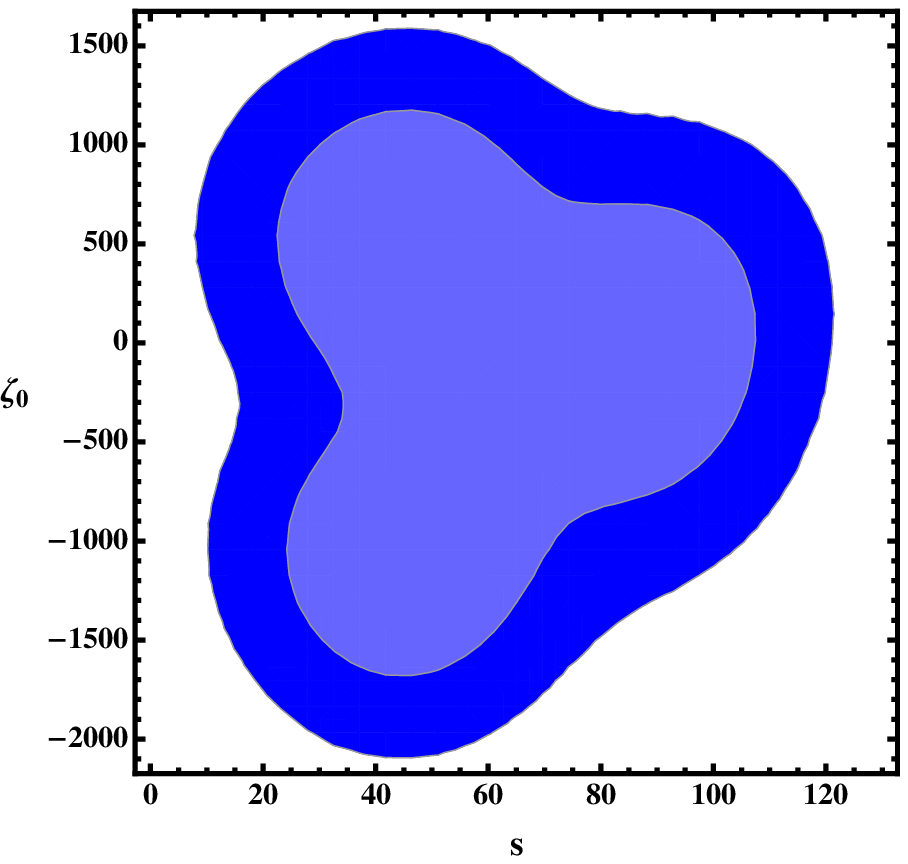}
\end{center}
\caption{Contour plots for the free parameters $\{\zeta_0, \Omega_m, s\}$ when fitting the Israel-Stewart model with data from Union 2.1. This case representes $\epsilon=0.1$.}
\label{Fig2} 
\end{figure*}

\begin{figure*} 
\begin{center}
\includegraphics[width=0.29\textwidth]{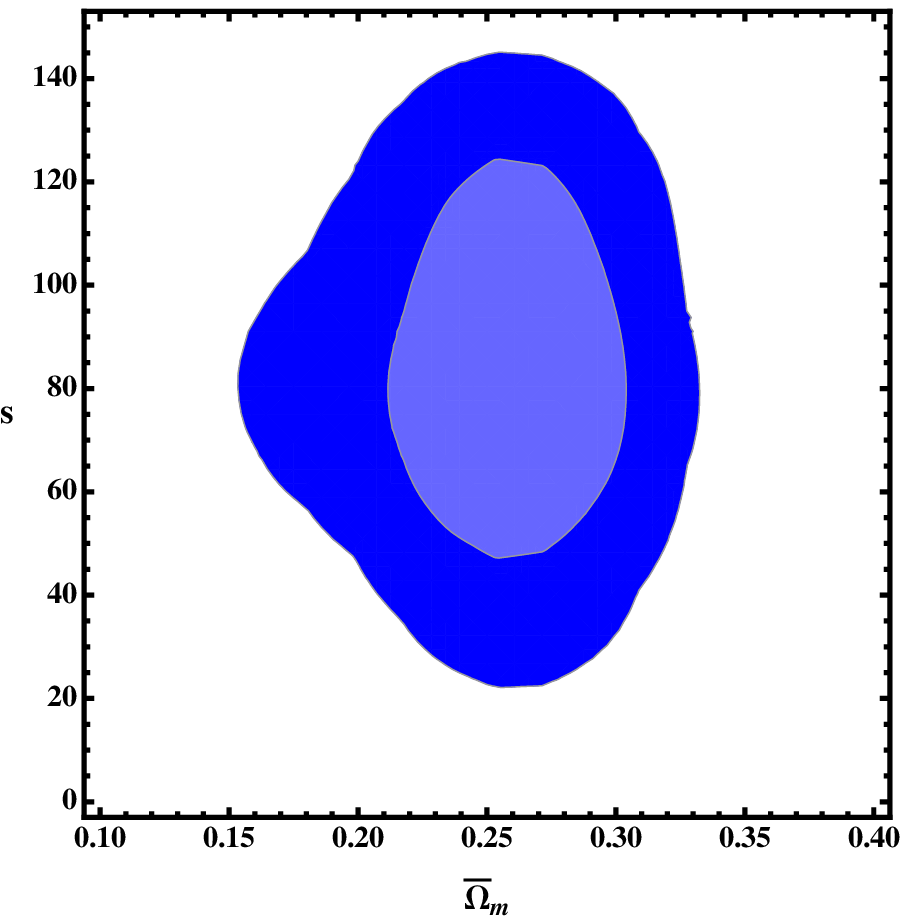}
\includegraphics[width=0.3\textwidth]{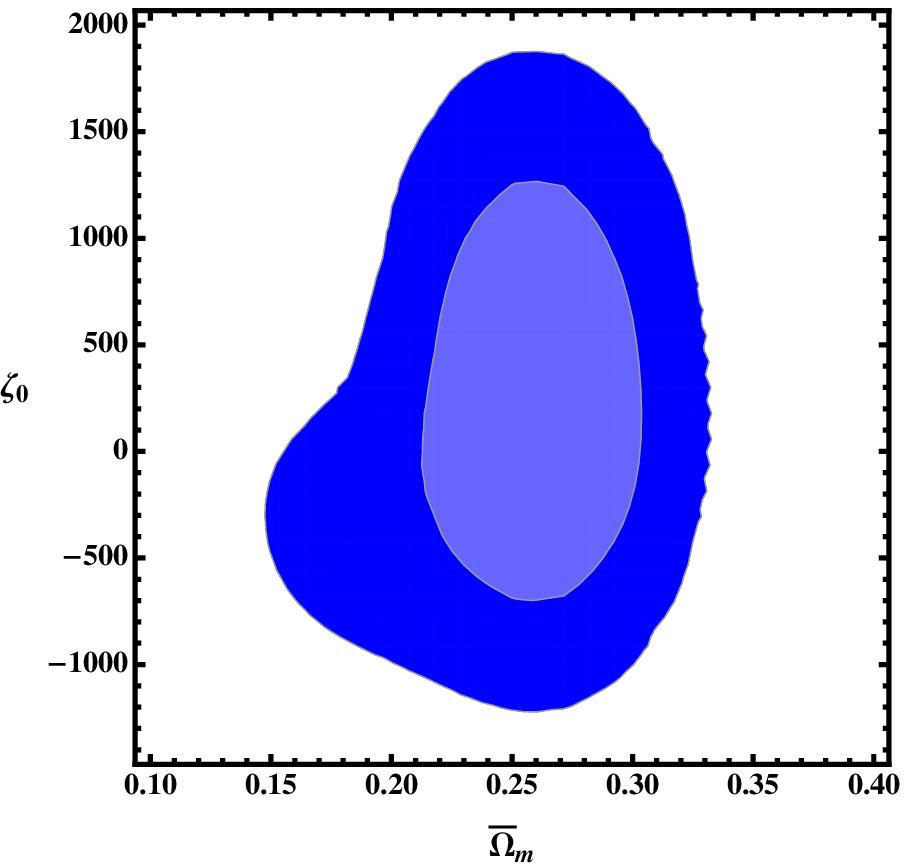}
\includegraphics[width=0.3\textwidth]{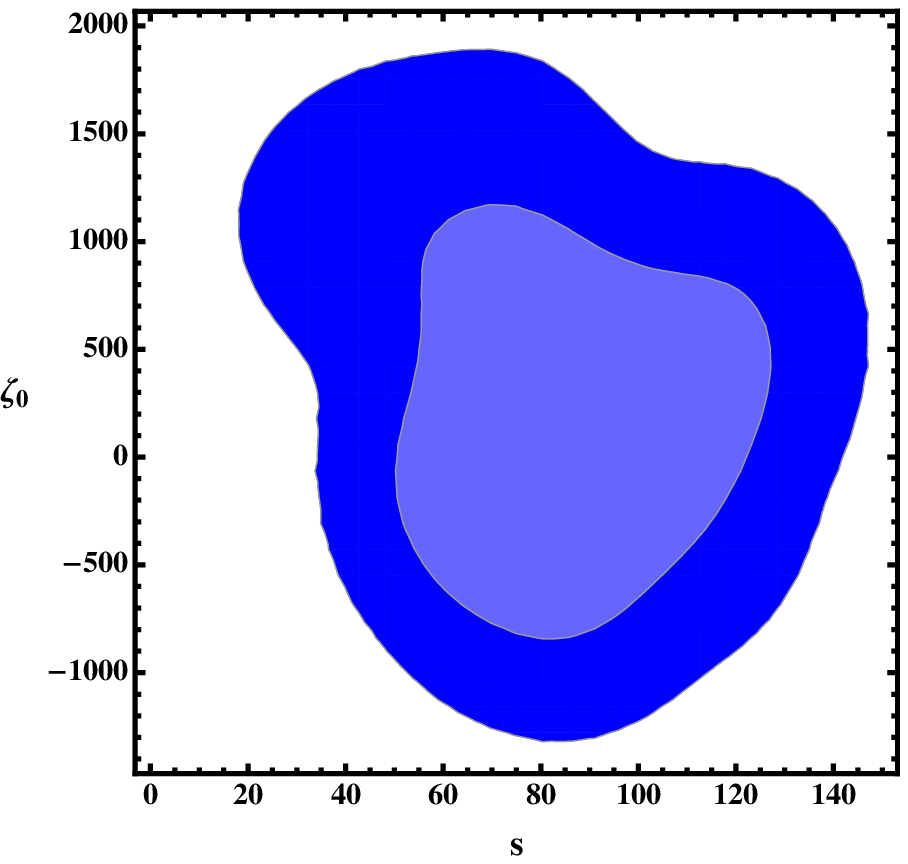}
\end{center}
\caption{Contour plots for the free parameters $\{\zeta_0, \Omega_m, s\}$ when fitting the Israel-Stewart model with data from Union 2.1. This case representes $\epsilon=0.01$.}
\label{Fig3} 
\end{figure*}
 
\section{Results and Discussion} \label{sect_results}

In order to simplify the equation (\ref{SN9}) and remove $H_0$, equation (\ref{SN9}) is solved by redefining the Hubble parameter as
\be
H(z)=H_0 E(z)\ ,
\label{Hubble}
\ee
where $H_0$ is the Hubble parameter evaluated today corresponding to the $\Lambda$CDM model while $E(z)$ is an adimensional function of the redshift. Then, by defining $\Omega_{\lambda}=\Lambda/3H_0^2$ and making use of the first FLRW equation (\ref{constraint}), the free parameter $\Omega_m$ enters in the equation as $\Omega_{\Lambda}=1-\Omega_m$. Moreover, we redefine the free parameter $\zeta_0$ to make it adimensional
\be
\zeta_0\rightarrow H_0^{1-2s} \zeta_0
\ee
In addition, the initial conditions are chosen to match $\Lambda$CDM model at a certain redshift $z_0$,
\be
E(z_0)=E^{\Lambda CDM}=\sqrt{\Omega_m (1+z_0)^3+1-\Omega_m}\ .
\label{InitialCond}
\ee
After solving the equation, the corresponding value of the $\Omega_m$ parameter has to be normalised by $E^2(0)$ in order to provide the real matter density predicted by the model, such that the matter density parameter is defined as follows
\be
\tilde{\Omega}_m=\frac{\Omega_m}{E^2(0)}\ .
\label{Omegam}
\ee
In the analysis we have imposed $\Omega_m>0$. The results are summarised in Table \ref{table1}, where we have also included the results for the $\Lambda$CDM model as a reference. In addition, Fig.~\ref{Fig1} and Fig.~\ref{Fig2} depict the counterplots for the free parameters for the cases $\epsilon=1$ and $\epsilon=0.01$. Moreover, it is meaningful to analyse the absolute constraining power of the SN data to this model by comparing the fits to the $\Lambda$CDM model. To do so, the goodness of fits is investigated by calculating the reduced $\chi^2$ values for each model and each case, which is is defined as follows
\be
\chi^2_{\rm red} =\dfrac{\chi^2_{\rm min}}{N-c-1}\ .
\ee
Here $N$ is the number of Supernovae considered from the Union 2.1 catalogue and $c$ is the number of free parameters of the model. In addition, we also calculate the so-called Bayesian complexity ($p_D$) and the Deviance Information Criterium (DIC) which provides additional information about a way of comparing different models and which are defined as follows
\be
p_{D}=\overline{\chi^2}-\chi^2_{\rm min}\ , \quad DIC=2\overline{\chi^2}-\chi^2_{\rm min}\ .
\ee
The $p_D$ parameter that depends on the deviation of the mean from the best fit provides a way to measure the goodness of the fits of a particular model compared to others. As shown in Table \ref{table1}, both $\Lambda$CDM model as the Israel-Stewart one (for any value of $\epsilon$) show a similar goodness. Also when analysing the DIC parameter, the conclusion shows up the same. Nevertheless, the values for $\chi^2_{red}$ are slightly smaller for the $\Lambda$CDM model due to the larger number of parameters of the Israel-Stewart model. Moreover, note that the parameter $\epsilon$ affects the mean and the corresponding errors of the free parameters. In particular, the larger $\epsilon$ is, the larger the errors for $\tilde{\Omega}_m$ are. This is due to the features of this particular model, since whether the speed of sound for the viscous fluctuations increases, the Israel-Stewart model itself is capable of reproducing late-time acceleration with no need of a cosmological constant, at the price of increasing the errors on $\tilde{\Omega}_m$ while the errors on $\{\zeta_0, s\}$ turns out a bit smaller. Nevertheless, one should expect a small $\epsilon$, such that the value of $\tilde{\Omega}_m$ as shown in Table \ref{table1} and Figs.~\ref{Fig2}-\ref{Fig3} matches better the usual value provided by $\Lambda$CDM model when several sources of data are used.


\begin{table*}
\begin{center}
\begin{tabular}{cccccc}
\hline
\hline
\bf{Model} & MCMC parameters   & $\bf{\chi_{\rm min}^2}$ & $\bf{\chi_{\rm red}^2}$ & $p_D$ & $DIC$\\
\hline \vspace{-5pt}\\
$\Lambda$CDM  & $\Omega_m= 0.27 \pm 0.02$\  & 542.683 & $0.97 $ & 1.0 & 544.734\\
\\
$\epsilon=1$ & $\tilde{\Omega}_m= 0.63 \pm 0.72$\ ,\  $\zeta_0=482\pm 342$\ ,\  $s=-1.47\pm 0.6$& $542.559$  & 0.98 & 1.4 & 545.384\\
\\
$\epsilon=0.5$ & $\tilde{\Omega}_m=0.32\pm0.13$\ ,\ $\zeta_0=579\pm 217$\ ,\  $s=-3.75\pm 1.52$ & $542.437$  & 0.98 & 1.2 & 544.927\\
\\
$\epsilon=0.1$ & $\tilde{\Omega}_m=0.26\pm0.02$\ ,\ $\zeta_0=-104.511\pm 649$\ ,\  $s=61\pm 20$ & $543.141$  & 0.98 & 1.0 & 545.138\\
\\
$\epsilon=0.01$ & $\tilde{\Omega}_m=0.26\pm0.03$\ ,\ $\zeta_0=284\pm 542$\ ,\  $s=85\pm 21$ & $542.671$  & 0.98 & 1.1 & 544.687\\

 \hline \hline
\end{tabular}
\end{center}
\caption{Mean values of the free parameters for the Israel-Stewart model for $\epsilon=1\ ,\ 0.1\ , \ 0.01$ and the $\Lambda$CDM model. The corresponding standard deviation is also shown. In addition, we also include the best fit $\chi^2_{min}$, the $\chi^2_{red}$, the Bayesian complexity and DIC values for the three cases.}
\label{table1}
\end{table*}
 

\section{Conclusions} \label{sect_conclusions}

In the present paper we have analysed the effects of considering a more realistic description for dark matter beyond the usual perfect fluid picture. To do so, we have followed the well-known Israel-Stewart approach, which describes the bulk viscosity of a particular fluid by adding just an effective pressure term in the continuity equation, in this case in the dark matter equation. By using the transportation equation for the viscous pressure, we have obtained the general equation that determines the evolution of the Hubble parameter, where a cosmological constant is included. The purpose of this paper, beyond those papers in the literature where bulk viscosity is considered to unify dark matter and dark energy, was based on the analysis of the effects of viscous dark matter with the presence of a cosmological constant that in principle is responsible solely for the late-time acceleration, in order to study possible exact solutions, particularly de Sitter solutions, the near equilibrium condition and how good the model fit to observational data.\\

By analysing equation (\ref{eqforH}), the existence of exact de Sitter solutions have been studied. Contrary to usual systems with the presence of perfect fluids, the presence of bulk viscosity admits exact de Sitter solutions, not only asymptotically as in the case of $\Lambda$CDM model. We have studied the wide range of solutions allowed by the equations,  depending on the parameters related to the bulk viscosity. In particular, the existence of such type of solutions provide some limitations on the parameter $\epsilon$, which recall that parametrises the speed of sound of non-adiabatic perturbations, playing an important role, as shown after when fitting the model with Sne Ia data. Also the possible solutions depending on the parameter $s$, which describes the bulk viscosity coefficient, are discussed and several solutions are obtained for different values of this parameter. Moreover, the condition to keep the fluid near the equilibrium is studied, where we found that the presence of the cosmological constant allows the viscous stress to be not necessarily greater than the equilibrium pressure $p$ in order to have an accelerating expansion.\\

It is necessary to point out that since during the fitting with the data, the parameter $s$ is kept free, there is no dependable way to ensure that the numerical solution of the Israel-Stewart equation will have an asymptotical behaviour close to the $\Lambda$CDM solution, or even the same behaviour when $\zeta_0$ goes to zero, as we have learned from the few type de Sitter solutions found. So, facing the lack of a complete understanding of possible solutions and their behaviour in a wide range of the parameter $s$, the constrains from the cosmological data are done as a preliminary investigation of the possible values of s that can accommodate in principle the data. Then, we have compared the model to observational data by using a catalogue of Sne Ia. We have performed several analysis depending on the speed of sound of viscous perturbations but keeping always a pressureless fluid ($w=0$). Despite the cases with $\epsilon$ close to 1 seem very unlikely as would provide a speed of sound close - or equal - to the speed of light, its analysis provides additional information about the behaviour and contribution to the acceleration of the expansion, which occurs when the speed of sound is large enough, a case studied previously in absence of cosmological constant in Ref.~\cite{Cruz2017}, where was shown that the expansion may even cross the phantom barrier for those cases. Nevertheless, such cases give larger errors in the matter density parameter, as shown in Table \ref{table1}. For more realistic values of $\epsilon$, the errors on $\tilde{\Omega}_m$ are similar to those of $\Lambda$CDM model but at the price of increasing the uncertainty on the parameters corresponding to the viscosity. Such large errors and values on the dissipative parameters seem to imply their irrelevance when comparing the model to data at small redshifts. In addition, note that the negativeness of $\zeta_0$ may imply violations of the second law of thermodynamics, despite we have not imposed a particular allowed range a priori during the realisations of the MCMC's. However, the goodness of the fits shows that the model is as good as $\Lambda$CDM model, but the presence and relevance of bulk viscosity remains uncertain.\\

Hence, we have performed a deep analysis of viscous dark matter by following Israel-Stewart formalism. In the future, additional analysis for studying the existence of other classes of cosmological solutions and the relevance of bulk viscosity in the growth of perturbations and the formation of large scale structure may provide additional information about the role of dissipative effects in the dark matter fluid, a task to be performed in the future.  

\section*{Acknowledgments}
N. C. acknowledges the support of CONICYT (Chile) through the grant Fondecyt $N^{\circ}$ 1140238. E.G. is supported by the CONICYT-PCHA/Doctorado Nacional/2016-21160331. D.S-C.G. is funded by the grant No.~IT956-16 (Basque Government, Spain) and by MINECO (Spain), project FIS2016-76363-P.

\end{document}